\documentclass[sigconf,nonacm]{acmart}
\AtBeginDocument{%
  }

\usepackage[ruled,vlined]{algorithm2e}

\copyrightyear{2025}
\acmYear{2025}
\setcopyright{acmlicensed}\acmConference[ICAIF '25]{6th ACM International Conference on AI in Finance}{November 15--18, 2025}{Singapore, Singapore}
\acmBooktitle{6th ACM International Conference on AI in Finance (ICAIF '25), November 15--18, 2025, Singapore, Singapore}
\acmDOI{10.1145/3768292.3770357}
\acmISBN{979-8-4007-2220-2/2025/11}

\newcommand{\cA}{\mathcal{A}}
\newcommand{\E}{\mathbb{E}}

\newcommand{\cL}{{\mathcal L}}
\newcommand{\cS}{{\mathcal S}}

\begin{document}

\title{Algorithmic pricing with independent learners and relative experience replay}


\author{Bingyan Han}
\orcid{0000-0003-0680-7967}
\email{bingyanhan@hkust-gz.edu.cn}
\affiliation{%
	\institution{Thrust of Financial Technology, The Hong Kong University of Science and Technology (Guangzhou)}
	\city{Guangzhou}
	\country{China}
}

\renewcommand{\shortauthors}{Han}

\begin{abstract}
 In an infinitely repeated general-sum pricing game, independent reinforcement learners may exhibit collusive behavior without any communication, raising concerns about algorithmic collusion. To better understand the learning dynamics, we incorporate agents' relative performance (RP) among competitors using experience replay (ER) techniques. Experimental results indicate that RP considerations play a critical role in long-run outcomes. Agents that are averse to underperformance converge to the Bertrand-Nash equilibrium, while those more tolerant of underperformance tend to charge supra-competitive prices. This finding also helps mitigate the overfitting issue in independent Q-learning. Additionally, the impact of relative ER varies with the number of agents and the choice of algorithms. 
\end{abstract}




\maketitle

\section{Introduction} \label{sec:intro}
Artificial intelligence has the potential to enhance the efficiency and quality of decision-making. In e-commerce, algorithmic pricing, where firms use computer algorithms to set product prices, has become increasingly prevalent. However, the sophistication of these algorithms raises serious concerns about collusion. Some worry that advanced pricing algorithms may learn that collusive strategies are optimal \cite{ezrachi2016virtual}. Although skepticism remains, recent experimental research suggests that dynamic pricing algorithms can independently develop collusive behavior without human oversight or direct communication. Empirical studies have also identified several cases indicative of potential collusion. For example, \cite{assad2020algorithmic} found that pricing algorithms used in German retail gasoline markets were indirectly associated with increased price-cost margins. Similarly, an Amazon seller was accused of price matching via algorithms \cite{miklos2019collusion}.

Algorithms can produce anticompetitive outcomes across various market environments \cite{hansen2020algorithmic,abada2023artificial,dou2025ai}. While firms may design algorithms to implement pre-arranged collusive agreements, this paper focuses on algorithmic collusion, the case where collusion arises solely through interactions among algorithms, without human involvement or communication \cite{waltman2008q,klein2019autonomous,calvano2020artificial}. Because firms often conceal their pricing strategies and high prices alone are difficult to assess legally, current antitrust enforcement relies on evidence of communication to detect collusion \cite{calvano2020protecting}. As a result, regulators may be unable to intervene if collusion emerges autonomously through algorithmic learning \cite{calvano2020protecting}. Since market efficiency depends on competition, recent studies \cite{calvano2020artificial,calvano2020protecting} have called for reforms to antitrust policy. These reforms have direct implications for firms using algorithmic pricing. It is therefore essential for policymakers, firms, and consumers to understand the learning dynamics underlying algorithmic collusion.

A key requirement for algorithmic collusion is that product pricing must be independent, which aligns naturally with the independent reinforcement learning (RL) paradigm. \cite{calvano2020artificial} documented the emergence of supra-competitive prices among autonomous pricing agents using independent tabular Q-learning (InTQL). They modeled the economic environment as a repeated Bertrand oligopoly game, which is essentially an iterated prisoner's dilemma. Firms supply differentiated products and compete by setting prices. Jointly setting high prices allows firms to earn monopoly-level profits. However, a firm can increase its profit by undercutting competitors and capturing more demand. The dilemma arises because if all firms undercut each other, they collectively earn significantly lower profits. In repeated games, agents can learn sophisticated strategies based on the history of play, allowing them to respond strategically to competitors' actions. This pricing game has been extensively studied in economics and computer science \cite{Sequential,wealthineq}. 

Two main limitations exist in the findings of \cite{calvano2020artificial}. First, from a technical standpoint, independent RL often suffers from overfitting to opponents' policies \cite{overfit}. Agents trained within the same instance may learn to cooperate, but those trained separately can behave differently and fail to sustain cooperation, even in identical environments. InTQL \cite{calvano2020artificial} exhibits this issue; firms trained in separate instances do not consistently charge supra-competitive prices together, which raises concerns about the real-world severity of algorithmic collusion.

Second, from an economic perspective, agents may differ in how they value relative performance (RP) in competition. In practice, many firms incorporate RP evaluation into executive compensation schemes to reduce the impact of common shocks and improve incentive alignment. The U.S. Securities and Exchange Commission (SEC, 2006) mandates disclosure of RP evaluation practices, including benchmarks and reference groups. Approximately 43\% of S\&P 500 firms employ RP-based incentives \cite{bizjak2016role}, supporting the assumption that some agents are averse to underperforming relative to peers. Nonetheless, not all agents share this concern; the analysis should also include agents indifferent to RP evaluation.

This paper integrates concerns about RP evaluation with experience replay (ER) \cite{lin1992self} in a multi-agent RL setting. Agents are assumed to sample experience tuples from a replay buffer using a softmax probability distribution. The inverse temperature parameter of this distribution reflects the agent's concern for RP and is referred to as the RP coefficient. A positive RP coefficient indicates tolerance toward underperformance relative to other agents, while a negative RP coefficient indicates aversion to underperformance and a preference for outperforming competitors. We refer to this framework as relative ER.

Our findings show that a positive RP coefficient tends to produce supra-competitive prices, whereas a negative RP coefficient leads to competitive outcomes consistent with the Bertrand-Nash equilibrium. The interpretation is intuitive: collusive pricing requires agents to be more tolerant of occasional defection during learning. In contrast, agents averse to underperformance prefer to charge lower prices to gain demand and increase profits, thereby converging to the Bertrand-Nash equilibrium. Notably, the proposed framework also mitigates the overfitting issue observed in InTQL. With a positive RP coefficient, even when agents trained in separate instances are paired to play against each other, the model still yields supra-competitive prices.

As in InTQL, we first use matrix-based approximations of optimal action-value functions to establish a baseline. We then validate our claims and examine robustness using deep Q-learning \cite{mnih2015human} with ER. The results are consistent, with one key distinction: deep Q-learning with uniform sampling converges to the Bertrand-Nash equilibrium. We further explore different numbers of agents and find that increased agent count makes collusion more difficult and reduces the likelihood of monopoly pricing. These findings suggest that while the RP coefficient plays a significant role, it is not the sole determinant of long-run outcomes. Achieving collusive behavior requires tuning the RP coefficient in conjunction with algorithmic design and the number of participating agents.

\subsection{Related work}  \label{sec:related}
Inequity aversion \cite{inequity} shares conceptual similarities with RP evaluation by incorporating reward disparities into the objective function. However, our proposed framework differs significantly from that of \cite{inequity}. First, \cite{inequity} applies relative rewards greedily without randomization. In contrast, the proposed relative ER framework samples experience tuples using a softmax distribution, effectively balancing greedy RP evaluation with exploratory sampling across transitions. This approach enhances robustness against noise spikes. Second, simple reward comparisons may be inappropriate for asymmetric games. To address this, we evaluate RP using price-cost margins, which provide a more reasonable basis under agent heterogeneity. Third, empirical results indicate that inequity aversion \cite{inequity} leads to supra-competitive prices but fails to achieve the Bertrand-Nash equilibrium. In contrast, relative ER is capable of learning both outcomes by adjusting the RP coefficient, offering greater flexibility.

Earlier work \cite{Wunder} demonstrated that Q-learning in multi-agent RL can converge to average payoffs exceeding the Nash equilibrium under certain conditions. However, their analysis focused on a repeated general-sum game with only two players and two actions. In comparison, this paper emphasizes methodology and empirical evaluation under more complex game settings. Moreover, the relative ER framework can be extended to other multi-agent RL problems, although convergence guarantees remain open due to the non-stationarity of multi-agent environments.

Experience replay is a foundational technique \cite{lin1992self} that was later popularized by deep Q-learning \cite{mnih2015human}. In single-agent settings \cite{mnih2015human}, ER with uniform sampling reduces correlations in observation sequences. Subsequent work \cite{priori} introduced stochastic prioritization based on temporal-difference (TD) errors. To address sparse rewards, hindsight ER \cite{hindsight} reinterpreted the attainment of specific states as alternative goals. \cite{fedus2020revisiting,DeeperLookER} examined ER hyperparameter configurations. In multi-agent RL, naive application of ER can result in instability or divergence. Several sampling techniques have been proposed to mitigate these issues; see \cite{stabilising} for an example. Recent ER variants include \cite{lu2023synthetic,bellitto2024saliency}, among others.

The remainder of the paper is organized as follows. Section \ref{sec:game} introduces the Bertrand oligopoly economic environment. Section \ref{sec:REP} describes the relative ER framework. Section \ref{sec:experiment} presents experimental results. Section \ref{sec:discuss} concludes the paper. Our code is publicly available at \url{https://github.com/hanbingyan/RelativeER}.

\section{Game formulation} \label{sec:game}
Consider an infinitely repeated pricing game with $n$ differentiated products and an outside good. Firms compete by setting prices for their respective products, with each firm owning exactly one product and adjusting prices simultaneously in each period. This setup corresponds to the Bertrand oligopoly model, a standard framework in the study of collusion. Assume that the demand $q_{i, t}$ for product $i$ in period $t$ follows the logit model:
\begin{equation}\label{Eq:Demand}
	q_{i, t} = \frac{e^{ \frac{b_i - a_{i, t}}{\mu}}}{ \sum^n_{j=1} e^{ \frac{b_j - a_{j, t}}{\mu}}  + e^{\frac{b_0}{\mu}}},
\end{equation}
where $a_{i,t}$ and $b_i$ are product price and quality index, respectively. $\mu$ is a constant describing horizontal differentiation between products. Products can perfectly substitute each other as $\mu \rightarrow 0$. Product $0$ is the outside good. $b_0$ is the inverse index for aggregate demand. For further details and motivation behind this specification, see \citet[Section II.A.]{calvano2020artificial}.

Consequently, reward for firm $i$ in period $t$ is 
$$
r_{i, t} = ( a_{i, t} - c_i ) q_{i, t}, 
$$
where $c_i$ denotes the constant marginal cost. Firm $i$ is more efficient than firm $j$ if $c_i < c_j$. The game is symmetric when all firms share the same $b_i, c_i$; otherwise, it is asymmetric. For simplicity, assume that all firms remain active throughout the repeated game. The terms firm, agent, and player are used interchangeably.

Consider a single-period game. When each firm maximizes its own profit independently, the resulting price, denoted by $a^N$, corresponds to the Bertrand-Nash equilibrium. In contrast, if all firms coordinate to maximize joint profits, the resulting price $a^M$ is higher and yields greater rewards. Although an individual firm can temporarily outperform its competitors by undercutting the price, mutual undercutting leads all firms to lower profits, which is a prisoner's dilemma.

In an infinitely repeated game, each firm $i$ aims to maximize its own discounted return:
\begin{equation}
	\E\Big[ \sum_{t = 0}^\infty \gamma^t r_{i, t} \Big],
\end{equation}
where $0 < \gamma < 1$ is a common discount factor.

Assume all firms share the same action space $\cA$, which includes all feasible prices. At each period $t = 0,1,2, \dots$, firms observe a common state $s_t \in \cS$, representing the current environment. For simplicity, we assume full observability of the state, excluding partial information settings. Firms select actions (i.e., prices) based on their policies $\pi_i$, which map the observed state space $\cS$ to the action space $\cA$. Both state and action spaces are assumed to be finite and discrete. While prices are publicly observable, policies remain private. Firms choose prices simultaneously, receive individual rewards, and the environment transitions to the next state $s_{t+1}$. We also assume that firms can observe their competitors' rewards. This assumption is reasonable, as public companies are typically required to disclose earnings regularly.

Monopolies suppress competition and reduce the efficiency of market systems. Consequently, most jurisdictions enforce antitrust laws to penalize monopolistic and collusive practices. Current regulations rely on evidence of communication to identify collusion \cite{calvano2020protecting}. However, these laws may not apply if firms learn collusive strategies through algorithms without explicit communication. This presents serious concerns regarding algorithmic collusion. To avoid communication, we adopt independent RL in this study. While this approach may not precisely replicate algorithms used in practice, it reflects the core dynamics of the iterated prisoner's dilemma and offers a tractable model for analysis. 

The optimal action-value function $Q^*_i(s, a_i)$ for firm $i$ is defined as the maximum expected payoff achievable by following any policy $\pi_i$ after observing state $s$ and taking an action $a_i \in \cA$:
\begin{equation}\label{Eq:def-Q}
	Q^*_i(s, a_i) = \max_{\pi_i} \E \Big[ \sum_{t = 0}^\infty \gamma^t r_{i, t} \Big| s, a_i, \pi_i \Big]. 
\end{equation}

A policy that achieves the maximum in \eqref{Eq:def-Q} is denoted by $\pi^*_i$. Firms are assumed to observe the actions and rewards of their competitors but not the internal policy structures, such as the functional forms or matrices representing those policies.

The function $Q^*_i(s, a_i)$ satisfies Bellman's equation:
\begin{equation}\label{Eq:Bellman}
	Q^*_i(s, a_i) = \E \Big[ r_i + \gamma \max_{a'_i \in \cA} Q^*_i(s', a'_i) \Big| s, a_i\Big],
\end{equation}
where $r_i$ is the one-period reward for agent $i$, and $s'$ is the next state observed after taking action $a_i$ under state $s$.

Independent learners treat the actions of other agents as part of the environment and employ either Q-matrices \cite{watkins1992q} or neural networks \cite{mnih2015human} to approximate $Q^*_i$. A trade-off exists between exploring suboptimal actions to gather new information and exploiting current knowledge. The $\varepsilon$-greedy policy suggests following the current greedy action with probability $1 - \varepsilon_t$ and a purely random action with probability $\varepsilon_t$. The exploration rate declines over time as follows:
\begin{equation}
	\varepsilon_t = e^{- \beta t},
\end{equation}
with $\beta > 0$.

To quantify profitability, we define the profit ratio as 
\begin{equation}\label{Eq:Ratio}
	\Delta = \frac{\bar r - r^N}{r^M - r^N}, 
\end{equation}
where $\bar{r}$ denotes the average reward per agent upon convergence, $r^N$ is the reward in the Bertrand-Nash equilibrium, and $r^M$ is the monopoly reward. A value of $\Delta = 0$ indicates competitive outcomes, while $\Delta = 1$ implies collusive behavior. The key question is whether independent learners will charge prices close to the competitive price $a^N$ or the monopoly price $a^M$. Using InTQL, profit ratios between 60\% and 90\% have been observed in \cite{calvano2020artificial}, substantially exceeding the competitive benchmark and illustrating the potential for algorithmic collusion.

\begin{figure}[h]
	\centering
	\includegraphics[width=0.8\linewidth]{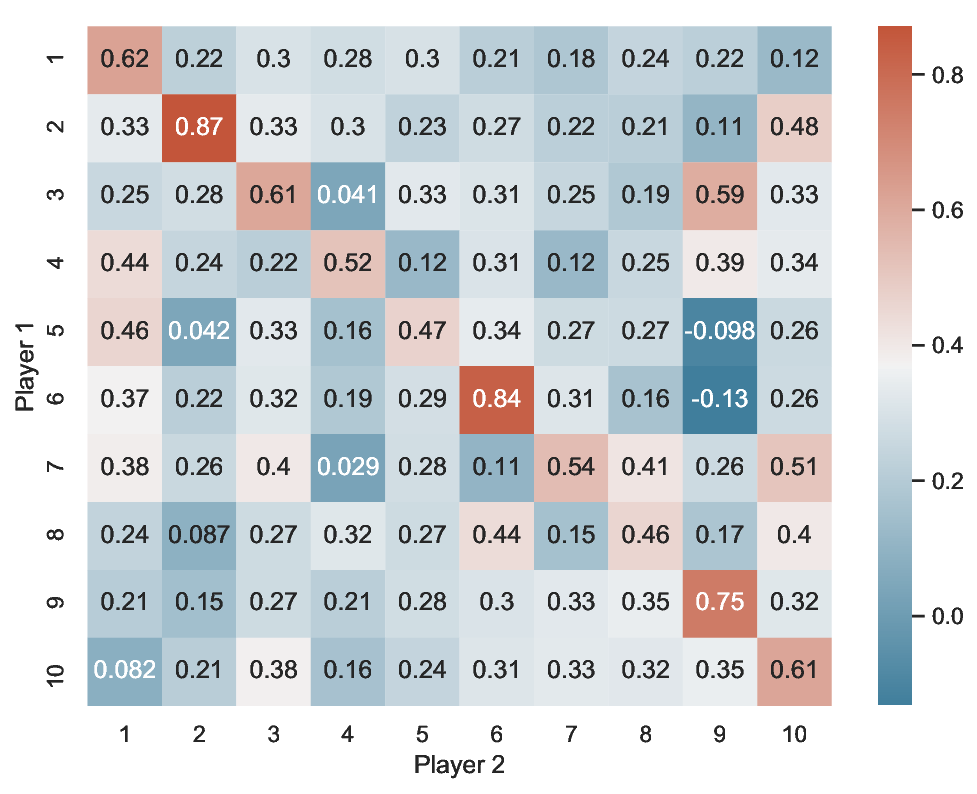}
	\caption{Profit ratios for independent tabular Q-learning (InTQL)} 
	\label{fig:aer}
	\Description{Ratio matrix for independent tabular Q-learning (InTQL)} 
\end{figure}

Despite this, InTQL \cite{calvano2020artificial} has two limitations. First, independent learning tends to overfit opponents' strategies and fails to generalize across different training instances \cite{overfit}. Upon convergence, both long-run prices and greedy policies derived from Q-functions are obtained. Consider a symmetric pricing game with two agents. Suppose $I$ instances of InTQL are trained separately. Selecting one agent from a particular instance and pairing it with an agent from another, we allow them to interact using their respective learned greedy strategies. Figure~\ref{fig:aer} displays the resulting profit ratios across 10 instances. Diagonal entries reflect agents trained in the same instance, while off-diagonal entries represent pairings from different instances. Profit ratios are consistently lower off-diagonal, confirming that cross-instance generalization is weak.

Second, agents may differ in their sensitivity to RP. Firms often use heterogeneous evaluation criteria, as discussed in Section \ref{sec:intro}. These behavioral differences can lead to distinct equilibrium outcomes. Underperformance can be measured by rankings based on rewards, but alternative metrics such as price-cost margins are also used. Section \ref{sec:REP} introduces a methodology that leverages ER \cite{lin1992self} to investigate this conjecture and mitigate overfitting.

\section{Relative experience replay}  \label{sec:REP}
A transition $(s_t, a_{i,t}, s_{t+1}, r_{i, t})$ forms an experience tuple for agent $i$. ER stores past transitions in a fixed-length replay buffer, from which a mini-batch is sampled during updates. By reusing experience tuples, ER improves sample efficiency, especially when acquiring new observations is costly. Modern ER was popularized by deep Q-networks \cite{mnih2015human} in single-agent RL. The standard approach samples transitions uniformly from the buffer, but this ignores differences in tuple significance. Prioritized ER \cite{priori}, which selects transitions with high TD errors, provides a more efficient and effective alternative.

In multi-agent RL, where opponents are treated as part of the environment, the problem becomes non-stationary from each agent's perspective. An agent's policy depends on those of its rivals, which evolve over time due to exploration. Consequently, older experience tuples may become outdated and fail to reflect current dynamics. This makes it important to prioritize transitions, for example by using sampling schemes that discount obsolete data in the replay buffer \cite{stabilising}.

To account for agents' heterogeneous attitudes toward RP, each agent $i$ maintains a matrix $D_i(s, a_i)$. After each period $t$, if agent $i$ is outperformed by $d_{i,t}$ competitors, the matrix is updated by
\begin{equation}\label{Eq:Dmatrix}
	D_i(s_t, a_{i,t}) \leftarrow D_i(s_t, a_{i,t}) + d_{i, t},
\end{equation} 
while all other entries remain unchanged. The corresponding experience tuple $(s_t, a_{i,t},$ $ s_{t+1}, r_{i, t})$ is stored with the label $D_i(s_t, a_{i,t})$, which serves as a measure of the transition's importance. The matrix $D_i$ is referred to as the RP matrix for agent $i$. 

Conceptually, $D_i(s, a_i)$ quantifies how severely agent $i$ was outperformed when choosing action $a_i$ in state $s$. Since agents visit states at varying frequencies, $D_i$ is updated unevenly and may be biased. However, accurate values across all state-action pairs are unnecessary for predicting underperformance. Replay buffers typically store only recent transitions, so accurate values are needed only in the recently visited region of $D_i$. Additionally, ranking entries in $D_i(s, a_i)$ further enhances robustness. 

A precise definition of RP is required to compute $d_{i, t}$, but this definition may vary across agents. In the pricing game, if rewards satisfy $r_i < r_j$, agent $i$ may interpret this as underperformance relative to agent $j$. Accordingly, $d_{i, t}$ can be defined as the number of agents earning higher rewards than agent $i$ in period $t$. However, this definition may be inadequate in asymmetric pricing games. For example, when a firm has a lower marginal cost $c_i$, it may obtain higher rewards even at equal or lower prices. In such cases, it is more appropriate to define RP based on the price-cost margin. Under this criterion, agent $i$ would consider $a_i - c_i > a_j - c_j$ as underperformance with respect to agent $j$. Section \ref{sec:experiment} explores both definitions in the asymmetric setting. 

Given the RP matrix, agent $i$ samples a transition $j$ from its replay memory $M_i$ with probability 
\begin{equation}\label{Eq:SampleProb}
	P(j) = \frac{e^{\lambda p_j}}{\sum_{k \in M_i} e^{ \lambda p_k}},
\end{equation}
where the summation is over all stored tuples in $M_i$, and $p_k$ denotes the priority for transition $k$. One approach is to set $p_k$ equal to the stored RP label. A more robust variant assigns $p_k = \textbf{rank}(k)$, where the rank is determined by sorting labels from smallest to largest, reducing sensitivity to outliers. The scalar $\lambda \in (-\infty, + \infty)$, referred to as the RP coefficient, quantifies the agent's sensitivity to RP. A positive $\lambda$ reflects tolerance to being outperformed. A zero value implies uniform sampling, reducing to classical ER \cite{mnih2015human}, indicating RP indifference. A negative $\lambda$ implies aversion to underperformance and a preference for transitions where the agent performed well.

Equation \eqref{Eq:SampleProb} draws inspiration from Boltzmann exploration, where the RP coefficient serves as the reciprocal of temperature. This sampling scheme offers several desirable properties. First, transition probabilities remain invariant under additive shifts to all entries in $D_i$. Second, stochastic sampling mitigates bias from $D_i$ during training and enhances exploration. Third, the RP coefficient provides a concise characterization of agent RP preferences and includes classical ER as a special case. Moreover, as noted by one of the reviewers, Equation \eqref{Eq:SampleProb} aligns conceptually with entropy-regularized Nash equilibria or quantal response equilibria \cite{SokotaDKLLMBK23}.

This approach is referred to as relative ER. The complete procedure is presented in Algorithm \ref{Algo:DQN}.

\begin{algorithm} 
	\SetAlgoLined
	\SetKwInOut{Input}{Input}
	\SetKwInOut{Output}{output}
	For each agent $i = 1, \dots , n$, initialize action-value function $Q_i$, replay memory $M_i$, and RP matrix $D_i$\\
	
	Initialize a random state $s_0$ \\
	\For{period $t = 0,..., T$}{
		For each agent $i$, choose action $a_{i,t}$ with $\varepsilon$-greedy policy\\
		Execute actions $(a_{1,t}, \dots , a_{n,t})$ simultaneously; each agent observes all rewards $r_{i,t}$ and actions \\
		Update $D_i$ with rule \eqref{Eq:Dmatrix} \\
		Move to the next state $s_{t+1}$ \\
		Store transition $(s_t, a_{i,t}, s_{t+1}, r_{i, t})$ into $M_i$ with label $D_i(s_t, a_{i,t})$ \\
		For each agent $i$, sample transitions from $M_i$ with probability  \eqref{Eq:SampleProb} \\
		For each agent $i$, update $Q_i$ with sampled mini-batches\\
	}
	\caption{Relative experience replay}\label{Algo:DQN}
\end{algorithm}

\section{Experiments}  \label{sec:experiment}
We begin with the baseline model, a symmetric pricing game introduced in~\cite{calvano2020artificial}. The setup includes two agents ($n = 2$), with marginal costs $c_i = 1$, quality indexes $b_i = 2$, $b_0 = 0$, differentiation parameter $\mu = 0.25$, and discount factor $\gamma = 0.95$. The Bertrand equilibrium price is approximately $a^N \approx 1.473$, while the monopoly price is approximately $a^M \approx 1.925$. The feasible price set is defined as $\cA = \{1.20, 1.24, \dots , 1.92, 1.96\}$, with equally spaced entries. Due to discretization, $a^N$ and $a^M$ are not exactly included in $\cA$. The state space is defined as the set of past actions from the last $m$ periods, and we assume $m = 1$. Our code is publicly available at \url{https://github.com/hanbingyan/RelativeER}.

We focus first on the two-agent case, since laboratory experiments have shown that tacit collusion is ``frequently observed with two sellers, rarely in markets with three sellers, and almost never in markets with four or more sellers'' \cite[p.17]{potters2013oligopoly}. In real-world markets, several documented price-fixing cases also involve a small number of firms. This is partly because monopolistic or collusive behavior often arises in markets dominated by a few major players. Our framework, however, accommodates an arbitrary number of agents. An extension to larger agent populations is discussed in Section \ref{sec:agents}.

\begin{figure}[h]
	\centering
	\includegraphics[width=0.8\linewidth]{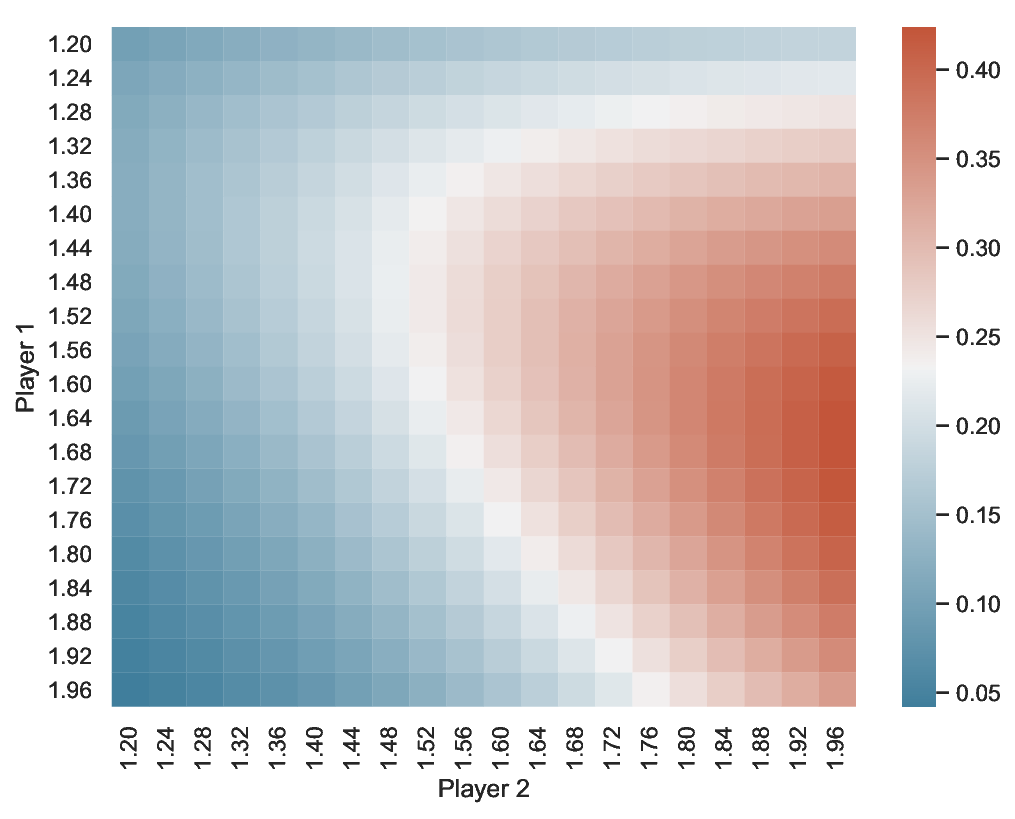}
	\caption{Reward for Player 1}
	\label{fig:reward}
	\Description{Reward for Player 1.}
\end{figure}

Figure \ref{fig:reward} displays the one-period reward for player 1 across all possible action pairs. Jointly charging supra-competitive prices yields higher rewards for both agents. However, a player may earn a greater individual reward by strategically undercutting the opponent. When both agents choose low prices, their rewards drop significantly. This outcome illustrates the prisoner's dilemma structure inherent in the pricing game. Under this symmetric setting, we assume that agents evaluate RP based on the realized rewards.

To illustrate how relative ER addresses the two issues discussed in Section \ref{sec:game}, we first examine the InTQL framework. The optimal action-value function $Q^*_i(s, a)$ is represented as an $|\cS| \times |\cA|$ matrix, given the finite state and action spaces. Classical Q-learning \cite{watkins1992q} initializes the action-value function with a matrix $Q_0$. At each time step $t$, agent $i$ selects an action $a_{i, t}$ under the current state $s_t$, receives a reward $r_{i, t}$ and observes the next state $s_{t+1}$. The algorithm then updates the matrix entry corresponding to $s = s_t$ and $a_i = a_{i, t}$ using the rule:
\begin{align}
	& Q_{i, t + 1} (s_t, a_{i, t}) \label{Eq:Qlearn} \\ 
	& = (1 - \alpha) Q_{i, t}(s_t, a_{i, t}) + \alpha \big[ r_{i, t} + \gamma \max_{a'_i \in \cA} Q_{i, t} (s_{t+1}, a'_i) \big], \nonumber
\end{align}
where all other entries remain unchanged. The learning rate $ \alpha \in [0, 1]$ is typically set to a small value to avoid discarding useful past information too quickly.

To incorporate relative ER into this framework, we modify the update rule in \eqref{Eq:Qlearn}. Instead of updating based solely on the most recent cell $(s_t, a_{i, t})$, we sample past transitions $(s_u, a_{i,u}, s_{u+1}, r_{i, u})$ for $ u \leq t$ from the replay buffer. The sampling probability is given by \eqref{Eq:SampleProb}, and the corresponding matrix entry is updated accordingly. In the special case where the buffer length is one and only one transition is sampled, the procedure reduces to classical Q-learning. Under the baseline setting, we set $\alpha = 0.15$. More generally, the learning rate is scaled inversely with the batch size for stability.

\begin{figure}[h]
	\centering
	\includegraphics[width=0.8\linewidth]{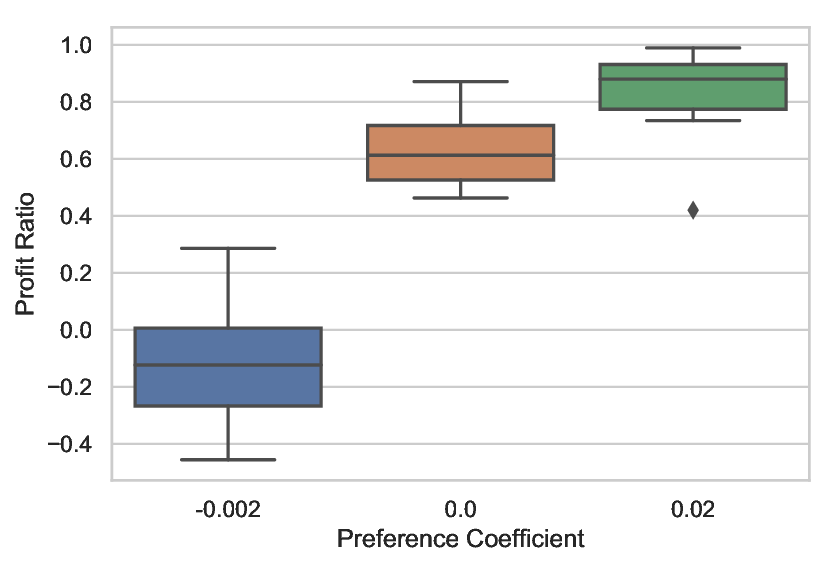}
	\caption{Profit ratios under InTQL}
	\label{fig:tabular_box}
	\Description{Profit ratios under InTQL}
\end{figure}

Figure \ref{fig:tabular_box} presents the distributions of profit ratios $\Delta$ from \eqref{Eq:Ratio} under varying RP coefficients, with all other hyperparameters held fixed. Given the relatively low dimensionality of the action and state spaces compared to typical RL applications, we use a memory buffer size of 1000, batch size of 8, and 10 instances per setting. For $\lambda = 0$, both the buffer and mini-batch sizes are set to 1. This case corresponds to the diagonal entries in Figure \ref{fig:aer}.

\begin{figure}[h]
	\centering
	\includegraphics[width=0.8\linewidth]{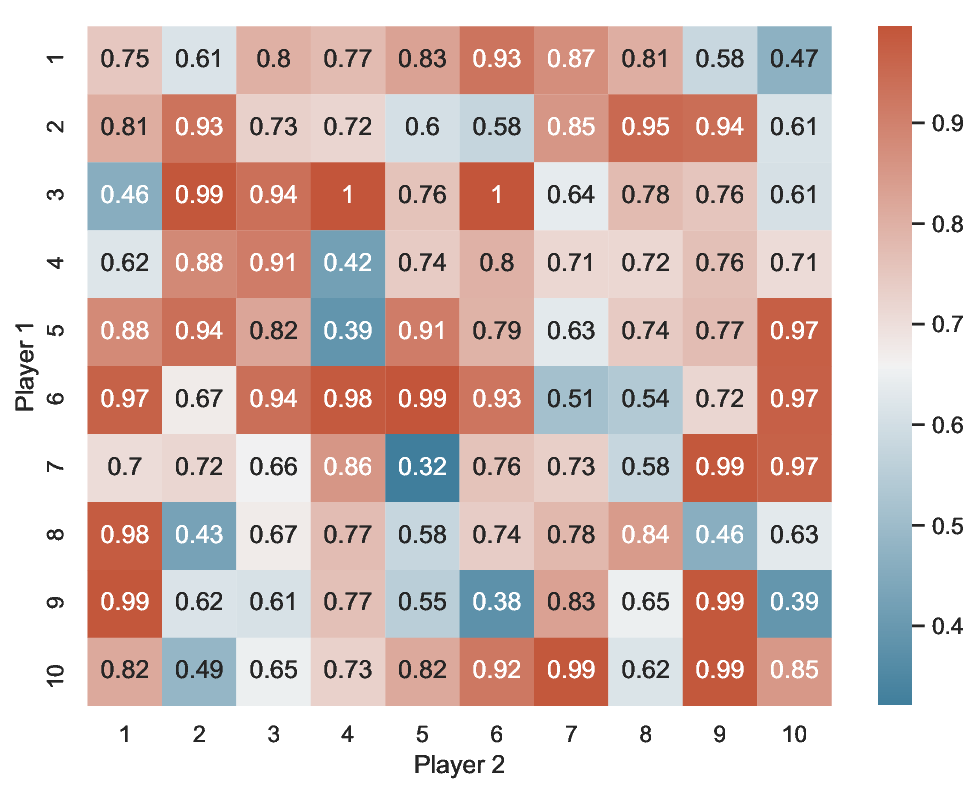}
	\caption{InTQL with tolerance avoids overfitting}
	\label{fig:tol_ratio}
	\Description{Profit ratios for InTQL with tolerance}
\end{figure}

The case $\lambda = 0.02$ reflects agents unconcerned with underperformance. These agents achieve significantly higher profit ratios compared to $\lambda = 0$. Figure \ref{fig:tol_ratio} shows that agents trained in separate instances also converge to supra-competitive prices near monopoly levels, indicating no evidence of overfitting. These agents are encouraged to replay experience tuples where competitors set lower prices, increasing exploration of supra-competitive prices across different states and instances. Relative ER also samples other low-price transitions randomly. The softmax sampling scheme promotes exploration beyond local optima, enhancing generalization and reducing overfitting.

In contrast, $\lambda = -0.002$ represents agents averse to underperformance. Most profit ratios fall below zero, meaning agents set prices lower than the Bertrand-Nash equilibrium. This behavior is expected, as lower prices carry a smaller risk of underperformance. The effect of $\lambda$ is asymmetric: a small negative value substantially lowers or eliminates algorithmic collusion, indicating that such anti-competitive outcomes are fragile.

The introduction of relative ER results in qualitatively different long-run behaviors depending on $\lambda$, supporting the hypothesis that agents' RP preferences critically shape outcomes. We proceed to assess robustness by varying the number of agents and incorporating deep Q-learning variants, revealing interactions between these factors and the RP coefficient.

\subsection{Noisy rewards}
A natural concern regarding RP is its robustness to observational noise. In previous analysis, the reward function $r_{i,t}$, derived from the demand function \eqref{Eq:Demand}, was assumed to be deterministic. However, in practice, observed profits may include random errors. This subsection considers the setting where agents observe rewards as $r_{i, t} + \varepsilon_{i, t}$, with $\varepsilon_{i, t} \sim N(0, \sigma^2)$ representing independent Gaussian noise. The objective is to assess the robustness of RP to such noise.

\begin{figure}[h]
	\centering
	\includegraphics[width=0.8\linewidth]{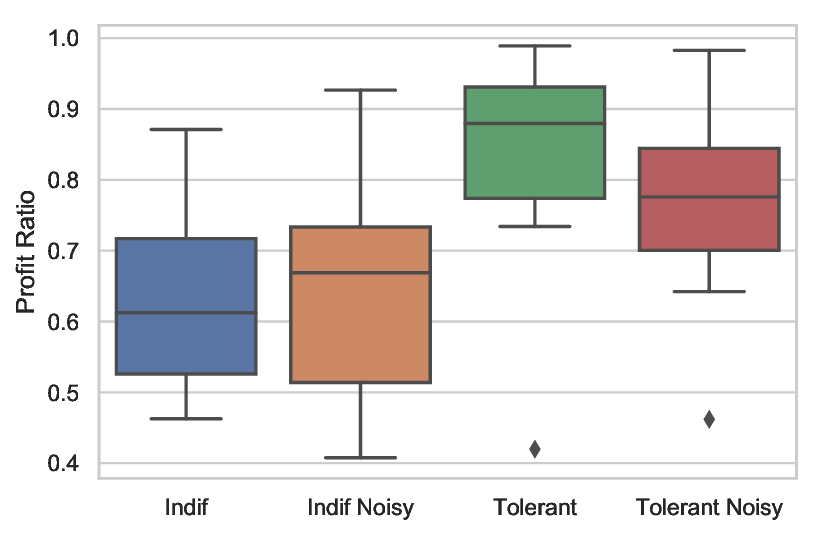}
	\caption{Profit ratios with noisy rewards}
	\label{fig:noisy}
	\Description{Profit ratios with noisy rewards}
\end{figure}

Figure \ref{fig:noisy} compares relative ER outcomes for $\lambda = 0$ and $\lambda = 0.02$, denoted as {\it Indif} and {\it Tolerant}, respectively. We assume the standard deviation $\sigma = 0.05$. Errors in observed rewards negatively impact collusion outcomes, and the results become more volatile. This occurs because relative ER with $\lambda > 0$ incorporates noisy rewards into the RP matrix computation, whereas the case with $\lambda=0$ uses rewards only in the $Q$-function update. As a result, the tolerant case is more sensitive to noise. While the median profit ratio decreases by approximately $0.1$, the outcomes still exceed those achieved when $\lambda = 0$. 

\subsection{Asymmetric games}
The previous sections considered only two agents with identical parameters. However, agents often differ in their capabilities, making the game asymmetric. This section focuses on cost asymmetry. Specifically, suppose agent 2 has a marginal cost of 0.5, making it more efficient than agent 1, whose marginal cost is 1. The action space is expanded to $\cA = \{1.20, 1.25, \dots , 2.25\}$ for both agents. The Bertrand-Nash equilibrium and monopoly prices are approximately $a^N \approx (1.372, 1.204)$ and $a^M \approx (2.198, 1.698)$. All other parameters follow the baseline setting. For simplicity, we report only relative ER with $\lambda = 0.1$.

Agent 2 earns higher profits even when charging the same price as agent 1. As shown in Table~\ref{tab:asy}, the converged price for agent 2 is generally higher than that of agent 1 when ranking transition importance based on rewards. Consequently, the total profit is not evenly distributed, as agent 1 is at a disadvantage. This outcome suggests that reward-based ranking is not appropriate in asymmetric settings.

A more reasonable criterion considers the price-cost margin, defined as $a_i - c_i$. If $a_1 - c_1 < a_2 - c_2$, then agent 1 outperforms agent 2. Table~\ref{tab:asy} shows that margin-based ranking leads to better outcomes than reward-based ranking. The average profit ratio is higher, and the standard deviation is smaller. Long-run average prices also indicate a fairer outcome for agent 1, who charges more than agent 2 in a greater proportion of periods.

One limitation of the margin-based approach is that it assumes agents have knowledge of their competitors' marginal costs. In practice, such information may not be publicly available. Therefore, asymmetry introduces additional challenges for sustaining collusion.

\begin{table}[t]
	\caption{Convergence results under asymmetric costs}
	\label{tab:asy}
	\begin{tabular}{c c c}\toprule
		Ranking criteria & Reward based & Margin based \\ \midrule
		Average $\Delta$ & 0.5290  & 0.6588 \\
		Standard deviation $\Delta$ & 0.1224 & 0.0826 \\
		Average prices & (1.7853, 1.8704) & (1.8162, 1.7341) \\
		Percentage of $a_{1} > a_{2}$  & 33.25\% &  53.34\% \\ \bottomrule
	\end{tabular}
\end{table}

\subsection{Number of agents}\label{sec:agents}

Collusion becomes more challenging as the number of agents increases. Consider a pricing game involving three agents. Under the same parameter settings as before, we have $a^N \approx 1.37$ and $ a^M \approx 2.0$. When the number of agents increases to six, these values become $a^N \approx 1.299$ and $ a^M \approx 2.132$. Table~\ref{tab:three} shows that three agents with $\lambda = 0.02$ achieve an average profit ratio close to zero. Although the average profit ratio improves with $\lambda = 0.1$, it remains much lower than in the two-agent case. To support higher pricing, agents must choose a larger $\lambda$ and be more tolerant of underperformance. In the six-agent case, achieving similar outcomes requires increasing $\lambda$ to $0.2$. 

\begin{table}[t]
	\caption{Comparison between two and three agents}
	\label{tab:three}
	\begin{tabular}{c c c}\toprule
		No. of Agents & Average $\Delta$ & Standard deviation $\Delta$ \\ \midrule
		2 ($\lambda=0.02$) & 0.8292  &  0.1572 \\
		3 ($\lambda=0.02$) & 0.0488 & 0.0795 \\
		3 ($\lambda=0.1$) & 0.5466 & 0.0757 \\  
		6 ($\lambda=0.2$) & 0.4681 & 0.0314 \\
		\bottomrule
	\end{tabular}
\end{table}

\subsection{Deep Q-learning}
Various RL algorithms exist, and we adopt the widely used deep Q-networks \cite{mnih2015human}, with a replay memory buffer size of 2000 and a mini-batch size of 16. Both agents use deep Q-learning with identical Q-network configurations: a fully connected neural network with a single hidden layer of size $h = 32$ and rectified linear unit (ReLU) activation.

Mini-batches sampled according to \eqref{Eq:SampleProb} are used to minimize the loss associated with the Bellman equation differences \cite{mnih2015human}. The optimal but unknown target values on the right-hand side of \eqref{Eq:Bellman}, given by $r_i + \gamma \max_{a'_i \in \cA} Q^*_i(s', a'_i)$, are replaced with the approximated values $r_i + \gamma \max_{a'_i \in \cA} Q_i(s', a'_i; \theta^-)$, where $\theta^-$ denotes network parameters from previous iterations. The Bellman error is defined as
\begin{equation*}
	\delta = Q_i(s, a_i; \theta) - \big[ r_i + \gamma \max_{a'_i \in \cA} Q_i(s', a'_i; \theta^-) \big].
\end{equation*}
Let $\cL(\delta; \theta)$ denote the loss function computed over a mini-batch of $\delta$ values sampled using relative ER. We use the squared error loss and optimize the parameters $\theta$ using the Adam optimizer with a learning rate of $10^{-4}$. After a fixed number of steps, the target network parameters $\theta^-$ are updated to match the current parameters $\theta$. Unlike supervised learning, the target values in deep Q-learning are not fixed and must be updated periodically \cite{mnih2015human}.

Thanks to small action and value spaces, deep Q-learning with relative ER converges in the Bertrand pricing game with approximately $10^6$ periods. Unlike InTQL with $\lambda = 0$, deep Q-learning with $\lambda = 0$ fluctuates among several prices near $a^N$. A possible explanation is that neural networks adapt more quickly than tabular Q-functions, making coordination and the discovery of monopoly prices more difficult, particularly under uniform sampling that disrupts temporal dependencies.

\begin{figure}[h]
	\centering
	\includegraphics[width=0.8\linewidth]{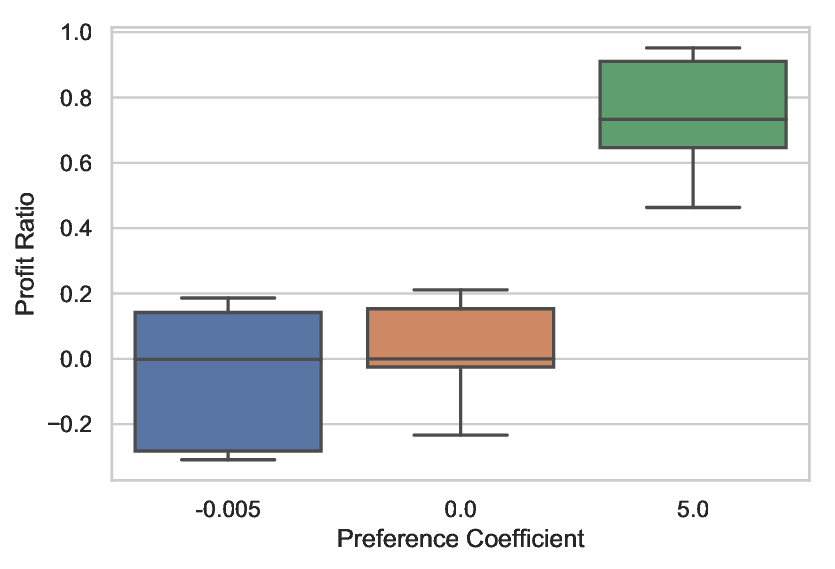}
	\caption{Profit ratios under deep Q-learning}
	\label{fig:dqn_box}
	\Description{Profit ratios under deep Q-learning}
\end{figure}

Relative ER in deep Q-learning produces similar effects to those observed in InTQL, while also exhibiting new behaviors. Figure \ref{fig:dqn_box} presents outcomes of deep Q-learning for various values of $\lambda$. Since deep Q-learning converges to the Bertrand-Nash equilibrium even when $\lambda = 0$, introducing a negative $\lambda$ does not enhance collusion and results in similar outcomes. However, overly small values of $\lambda$ often cause divergence and fail to reduce profit ratios. Achieving profit ratios $\Delta$ close to one requires a substantially larger RP coefficient.

\begin{figure}[h]
	\centering
	\includegraphics[width=0.8\linewidth]{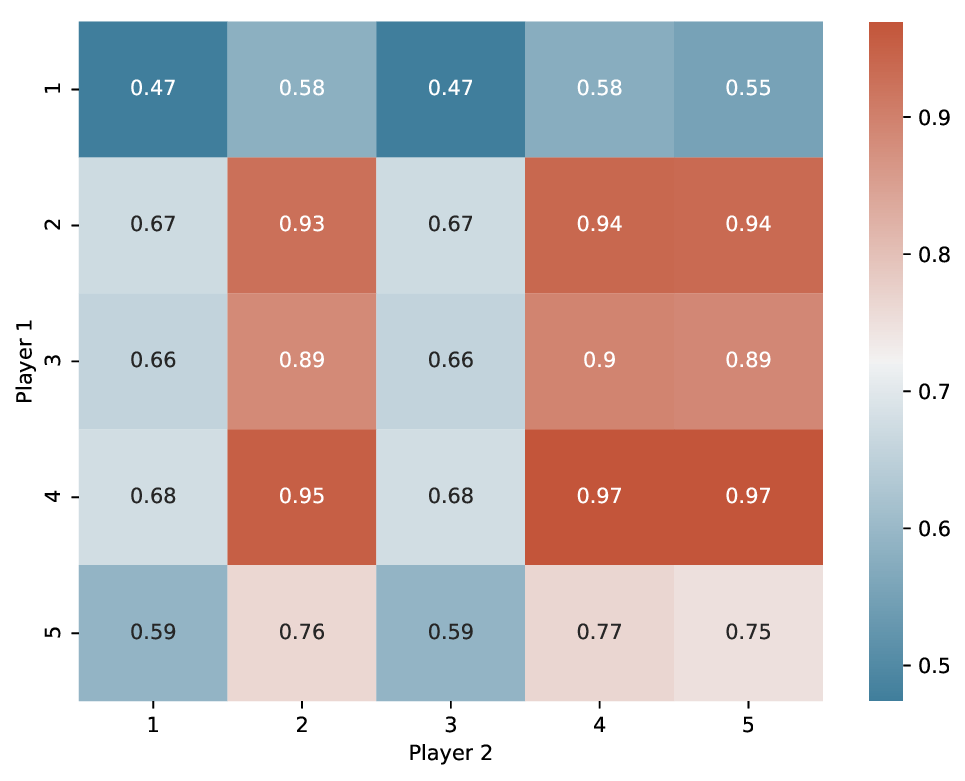}
	\caption{Deep Q-learning with tolerance avoids overfitting}
	\label{fig:dqn_ratio}
	\Description{Deep Q-learning with tolerance avoids overfitting}
\end{figure}

Another notable finding is that deep Q-learning with a positive $\lambda$ converges significantly faster than under uniform sampling, with the required number of periods reduced by nearly half. This result is important, as faster convergence increases the plausibility of algorithmic collusion in more realistic economic environments. In addition, Figure \ref{fig:dqn_ratio} shows that deep Q-learning with $\lambda = 5.0$ effectively avoids overfitting. Results are reported for five independent instances. 

\section{Discussion} \label{sec:discuss}

This paper introduces relative ER by incorporating RP concerns into multi-agent RL. Numerical results show that agents averse to underperformance adopt Bertrand-Nash equilibrium prices, while those tolerant of underperformance charge supra-competitive prices. These findings highlight RP as a key factor in shaping outcomes in general-sum pricing games. Several directions remain open for future investigation.

The current formulation assumes that rewards are observable at each time step, allowing efficient implementation of relative ER. However, in environments with sparse or delayed rewards, where feedback is infrequent, relative ER may become inefficient or even ineffective. A promising direction for future work is to explore how RP concerns can be addressed in such settings.

The analysis also assumes that both agents use the same learning algorithm. An important open question is whether agents using heterogeneous algorithms can still learn to coordinate in the absence of communication. Preliminary results suggest that such cases exhibit greater volatility and less consistent long-term behavior across runs.

Finally, this study adopts an experimental perspective to examine algorithmic collusion. A central theoretical challenge remains: to establish convergence guarantees for collusive or competitive strategies, along with rates of convergence, in multi-agent learning environments. Some recent advances can be found in \cite{cartea2022algorithmic,cont2024dynamics,bertrand2025self}. A relevant literature is Markov potential game \cite{LeonardosOPP22,FoxMOP22}.

\begin{acks}
Bingyan Han is partially supported by the HKUST (Guangzhou) Start-up Fund G0101000197 and the Guangzhou-HKUST(GZ) Joint Funding Program (No. 2024A03J0630). The author expresses gratitude to the anonymous referees for their valuable comments and suggestions that have greatly improved this manuscript. An earlier version of this paper was circulated and cited under the title ``Understanding algorithmic collusion with experience replay''.
\end{acks}



\end{document}